\documentclass[a4paper]{jpconf}
\usepackage{graphicx}
\usepackage{subfigure}

\begin{document}
\title{Chemical and thermal freeze-out of identified hadrons at the LHC}

\author{Xiangrong Zhu$^{1}$, Huichao Song$^{1, 2, 3}$}
\address{$^{1}$Department of Physics and State Key Laboratory of Nuclear Physics and Technology, Peking
University, Beijing 100871, China }
\address{$^{2}$Collaborative Innovation Center of Quantum Matter, Beijing 100871, China}
\address{$^{3}$Center for High Energy Physics, Peking University, Beijing
100871, China}

\begin{abstract}
This proceeding briefly summarizes our recent {\tt VISHNU} hybrid model investigations on the chemical and thermal freeze-out of various hadrons species in 2.76 A TeV Pb+Pb collisions. Detailed analysis on the evolution of particle yields and the last elastic collisions distributions during the hadronic evolution reveals that the two multi-strange hadrons, $\Xi$ and $\Omega$, experience early chemical and thermal freeze-out when compared with other hadron species.
\end{abstract}

\section{Introduction\label{sec:intro}}
\quad In relativistic heavy-ion collisions at top RHIC and the LHC energies, a hot and dense matter -- the Quark-Gluon Plasma (QGP) has been created. During its expansion, the QGP fireball quickly cools down, which then undergos the phase transition and produces a large amount of hadrons. These hadrons subsequently experience frequent inelastic and elastic collisions during the hadronic evolution. With the termination of inelastic and elastic collisions, the evolving system reach the chemical and thermal freeze-out, respectively.

Traditionally, the chemical freeze-out temperature $T_{ch}$ and the baryon chemical potential $\mu_{b}$ are extracted from the particle yields of various hadrons using the statistical model~\cite{BraunMunzinger:2001ip,Andronic:2005yp,Becattini:2005xt}, which gives a uniform $T_{ch}$
and $\mu_{b}$ for all hadrons species. Similarly, pure hydrodynamic simulations implement a uniform thermal freeze-out temperature to construct the freeze-out hyper-surface, which decouples various hadrons from the bulk matter~\cite{Peter}.

Microscopically, the related inelastic and elastic scattering channels are different for each hadron species. As a result, the chemical and thermal freeze-out procedures are very possibly hadron-species dependent. In this proceeding, we will briefly summarize our recent investigation on the chemical and thermal freeze-out of various hadrons in 2.76 A TeV Pb+Pb collisions~\cite{SongMultistrange}, based on dynamical simulations from {\tt VISHNU} hybrid model.

\begin{figure}[t]
  \centering
    \begin{minipage}[c]{.6\textwidth}
    \centering
    \includegraphics[width=0.8\textwidth]{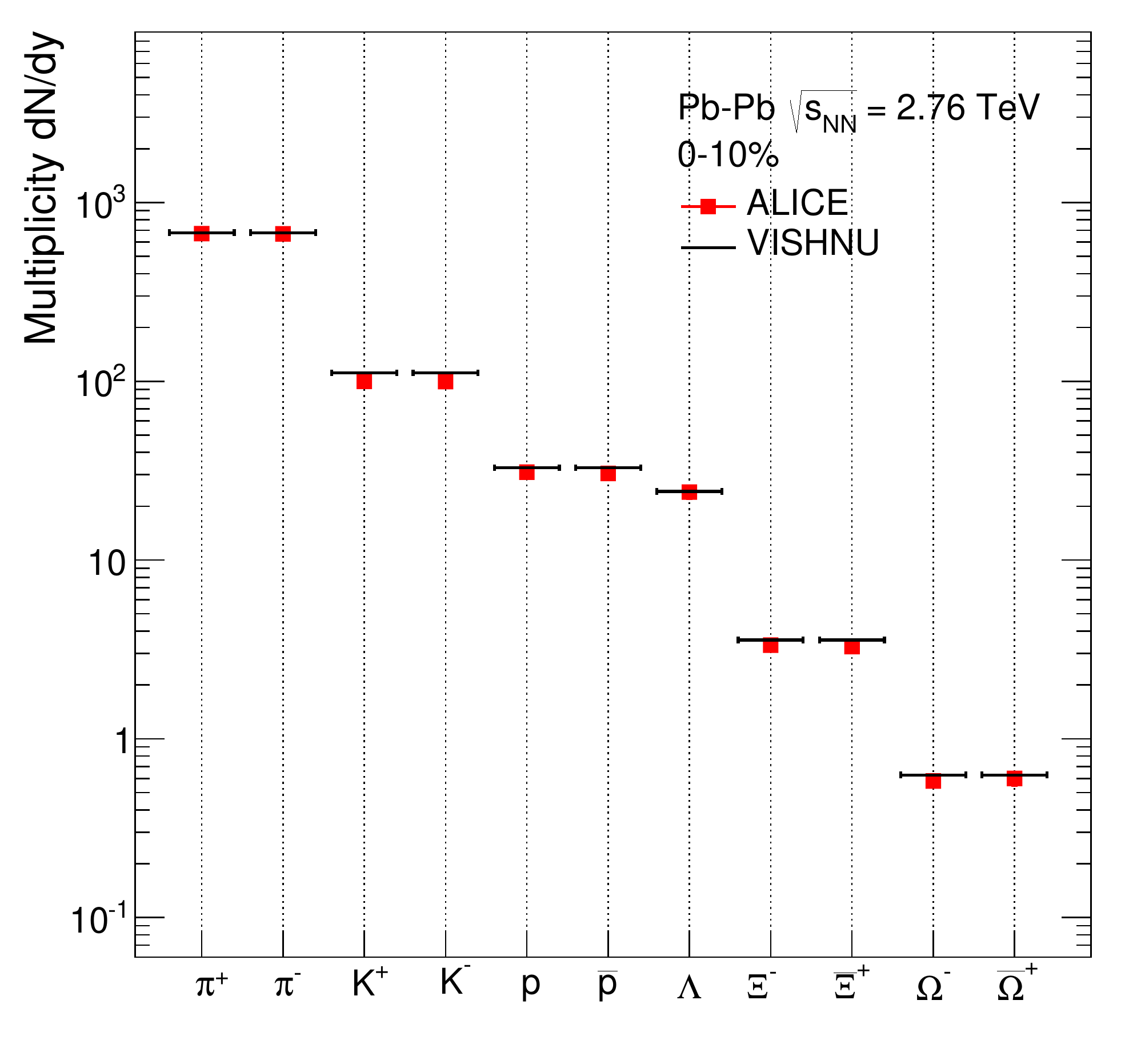}
  \end{minipage}
  \hspace{-0.2cm}
  \begin{minipage}[c]{.35\textwidth}
    \centering
    \caption{(Color online)Particle yields of $\pi$, $K$, $p$, $\Lambda$, $\Xi$, and $\Omega$ in the most central Pb+Pb collisions at $\sqrt{s_{NN}}$=2.76 TeV. The experimental data are from the ALICE Collaboration~\cite{Abelev:2013vea, Abelev:2013xaa, ABELEV:2013zaa}.
    Theoretical curves are from {\tt VISHNU} hybrid model calculations, using {\tt MC-KLN} initial conditions, $\eta/s=0.16$ and $T_{sw}=165 \ \mathrm{MeV}$.}
    \label{fig:VISHNUFit}
  \end{minipage}%
\end{figure}

\section{Setup of the calculations\label{sec:setup}}

\quad Our calculations implement {\tt VISHNU} hybrid model~\cite{Song:2010aq} that combines (2+1)-d relativistic viscous hydrodynamics ({\tt VISH2+1})~\cite{Song:2007fn} for the QGP fluid expansion with a microscopic hadronic transport model ({\tt UrQMD})~\cite{Bass:1998ca} for the hadron resonance gas evolution. The hydrodynamic calculations input the equation of state (EoS) s95p-PCE~\cite{Huovinen:2009yb} with the transition to {\tt UrQMD} at a constant temperature of 165 MeV. Following Ref.~\cite{Song:2011qa,Song:2013qma}, we input smooth initial conditions generated from the MC-KLN model~\cite{Drescher:2006ca,Hirano:2009ah}, where the normalization factor of the initial entropy density profiles is tuned to reproduce the multiplicity of all charged hadrons in the most central Pb+Pb collisions~\cite{Aamodt:2010cz}. The hydrodynamic simulations start at $\tau_0=0.9~{\rm fm}/c$ with the specific shear viscosity $(\eta/s)_{QGP}$ set to 0.16, and the bulk viscosity set to zero~\cite{Song:2011qa,SongQM}.

\section{Results}

\quad The chemical and thermal freeze-out of the evolving system are respectively related to the terminations of inelastic and elastic collisions. After the chemical freeze-out, the hadron yields no longer change. After the thermal freeze-out, the momentum distributions of various hadrons no longer change. In this section, we will briefly summarize our recent investigations on the chemical and thermal freeze-out of various hadron species at the LHC, based on dynamical simulations from {\tt VISHNU} hybrid model.

Figure~1 presents the {\tt VISHNU} calculations of the particle yields for $\pi$, $K$, $p$, $\Lambda$, $\Xi$, and $\Omega$ in the most central Pb+Pb collisions, which nicely fit the corresponding experimental data measured by ALICE. It was found that the baryon and anti-baryon ($B$-$\bar{B}$) annihilations during the hadronic evolution reduce the proton and anti-proton ($p/\bar{p}$) yields by $\sim30\%$, largely improving  the description of the $p/\bar{p}$ data~\cite{Song:2013qma}. Ref.~\cite{Song:2013qma} and~\cite{SongMultistrange} also showed that {\tt VISHNU} hybrid model also nicely fit the multiplicities,  $p_T$ spectra and differential elliptic flow of $\pi$, $K$, $p$, $\Lambda$, $\Xi$, and $\Omega$ at various centrality bins. With the good descriptions of these related soft hadron data, it is meaningful to further investigate the chemical and thermal freeze-out procedures of various hadrons at the LHC.

\begin{figure}[t]
  \centering
    \begin{minipage}[c]{.6\textwidth}
    \centering
    \includegraphics[width=0.98\textwidth, height=9cm]{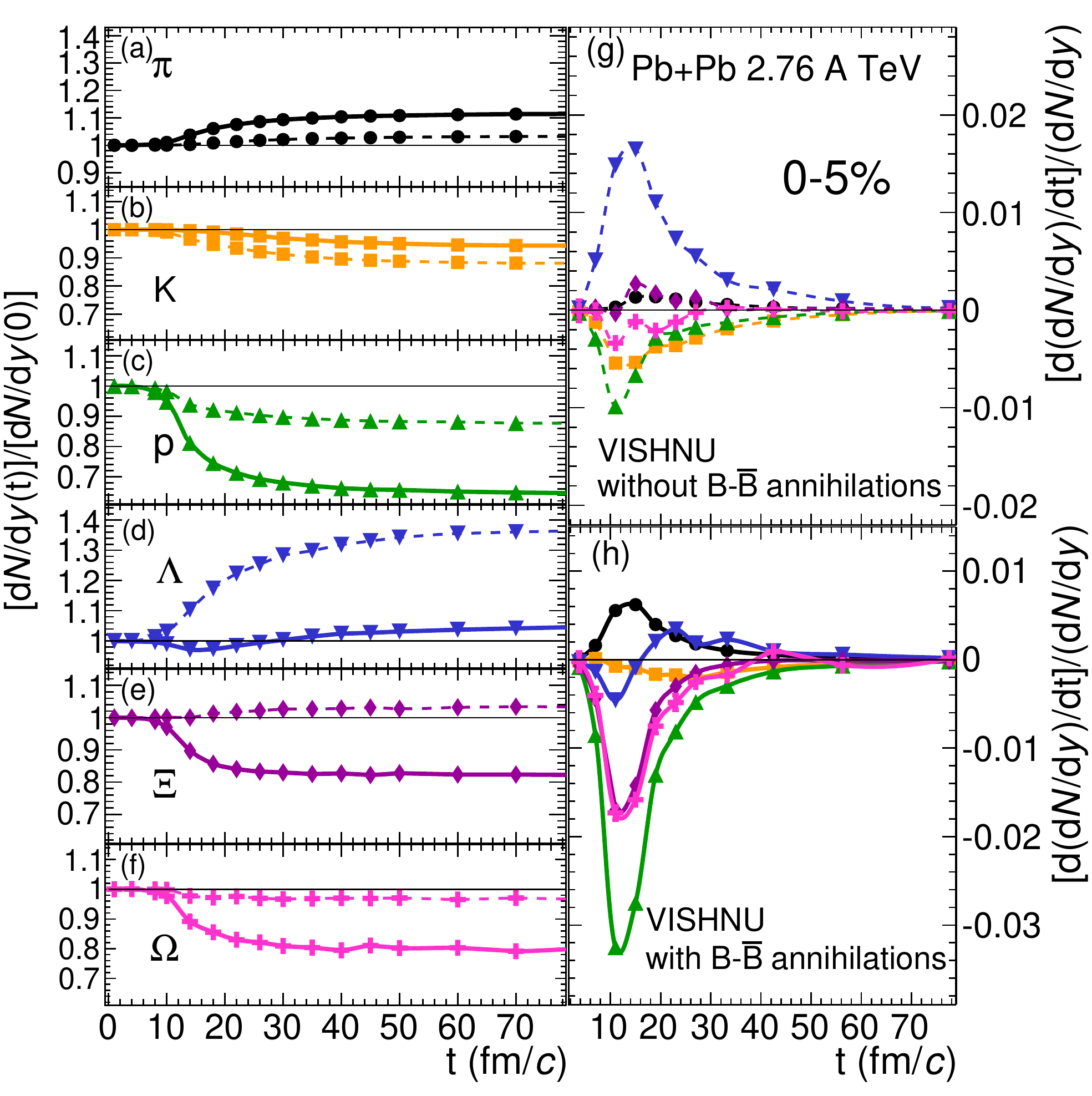}
  \end{minipage}
  \begin{minipage}[c]{.35\textwidth}
    \centering
    \caption{(Color online) Left panels (a)-(f): time evolution of relative particle yield density
    $\frac{dN}{dy}(t)/\frac{dN}{dy}(0)$ for $\pi$, $K$, $p$, $\Lambda$, $\Xi$, and $\Omega$ during the {\tt UrQMD} expansion of {\tt VISHNU}. Here, $\frac{dN}{dy}(t)$ and $\frac{dN}{dy}(0)$ denote the particle yield density at mid-rapidity at later evolution time and at the starting time, respectively.
Right panels (g) and (h): time evolution of the changing rates for the corresponding particle yield densities.
    Solid/dashed lines denote the {\tt VISHNU} simulations with/without ${\it B}$-$\bar{B}$ annihilations.}
    \label{fig:timedNdy}
  \end{minipage}
\end{figure}

Figure~\ref{fig:timedNdy} shows the time evolution for the particle yields of $\pi$, $K$, $p$, $\Lambda$, $\Xi$ and $\Omega$ during the hadronic evolution, simulated from {\tt VISHNU} hybrid model with and without ${\it B}$-$\bar{B}$ annihilations. For the simulations without ${\it B}$-$\bar{B}$ annihilations, the yields of these two multi-strange baryons $\Xi$ and $\Omega$ almost do not change. We also notice that their yields quickly reach saturation for the case with ${\it B}$-$\bar{B}$ annihilations. This reveals that $\Xi$ and $\Omega$ experience early chemical freeze-out after the hadronazition. In contrast, the yields of $p$ slowly decreases during the hadronic evolution especially for the case with ${\it B}$-$\bar{B}$ annihilations, indicating later chemical freeze-out of protons. On the production of $\Lambda$ and $K$, the ${\it B}$-$\bar{B}$ annihilations almost balance with other inelastic collision channels.  As a result, the yields of $\Lambda$ and $K$ only slightly change during the hadronic evolution with ${\it B}$-$\bar{B}$ annihilations~\cite{SongMultistrange}.

The right panels of Fig.~1 presents the changing rates of the particle yields of these identified hadrons. For the simulations without ${\it B}$-$\bar{B}$ annihilations, the changing rates for $K$, $p$, and $\Lambda$ show wide peaks along the time axis, illustrating that the associated inelastic collisions are still frequent after 10-20 ${\rm fm}/c$. Due to small hadronic cross sections, multi-strange hadrons $\Xi$ and $\Omega$ rarely interact with other hadrons. While, the annihilations with their own anti-particles lead to wide peaks of the changing rates for $\Xi$ and $\Omega$ in panel (h), which
delay their chemical freeze-out, when compared with the case without ${\it B}$-$\bar{B}$ annihilations. However, these multi-strange hadrons still experience earlier  chemical freeze-out in the full {\tt VISHNU} simulations compared with other identified hadrons~\cite{SongMultistrange}.

\begin{figure}[t]
  \centering
    \begin{minipage}[c]{0.6\textwidth}
    \centering
    \includegraphics[width=0.8\textwidth, height=7.0cm]{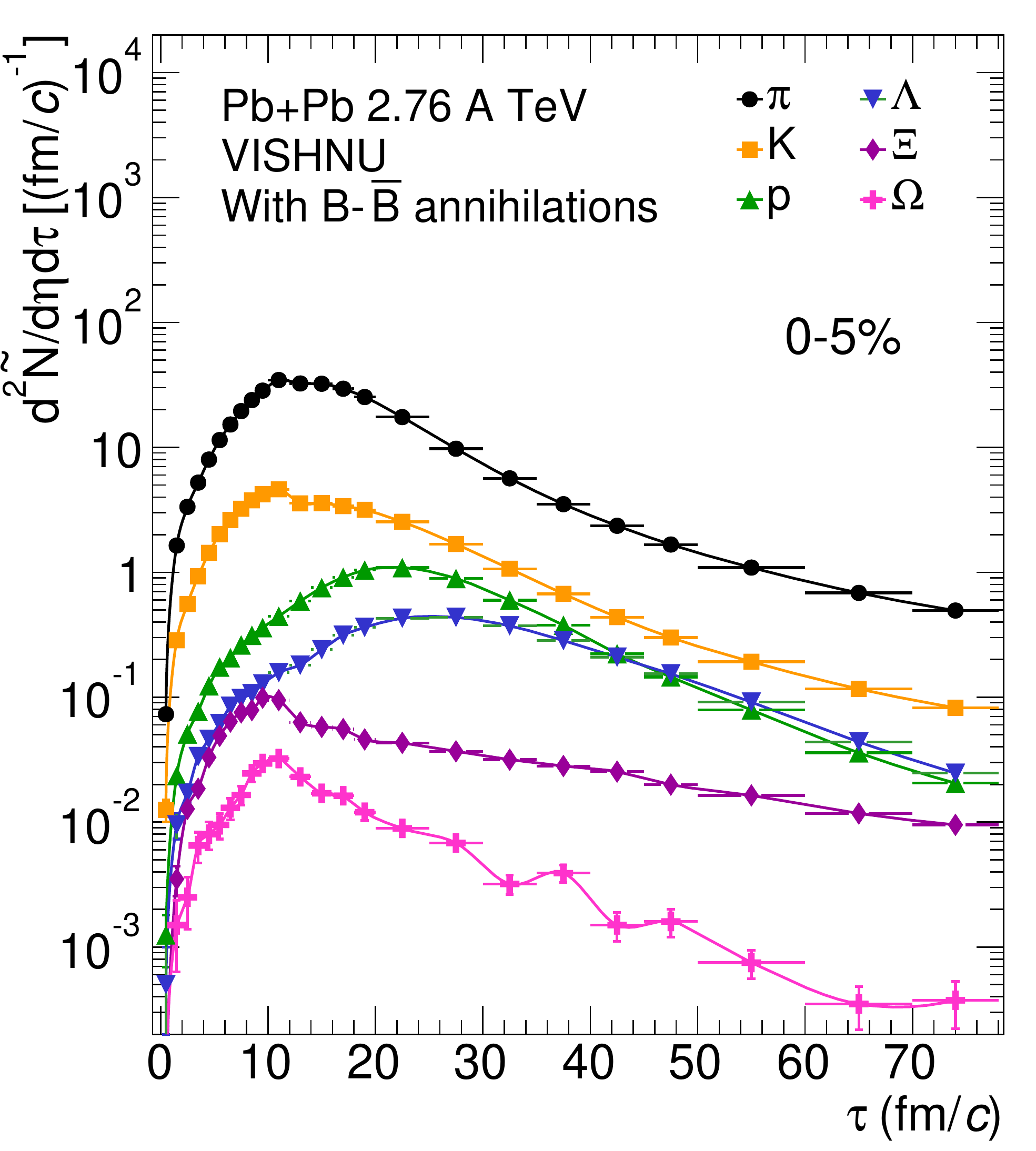}
  \end{minipage}
  \begin{minipage}[c]{.3\textwidth}
    \centering
    \caption{(Color online) Thermal freeze-out time distributions for $\pi$, K, p, $\Lambda$, $\Xi$ and $\Omega$ in the most central
   Pb+Pb collisions, calculated from {\tt VISHNU} with ${\it B}$-$\bar{B}$ annihilations.}
    \label{fig:WBB05}
  \end{minipage}%
\end{figure}

Figure~\ref{fig:WBB05} presents the thermal freeze-out time distributions of $\pi$, $K$, $p$, $\Lambda$, $\Xi$ and $\Omega$ in the most central Pb+Pb collisions. Here, the thermal freeze-out time distributions are extracted through analyzing the space-time information of the last collisions for various hadron species.  For $\Xi$ and $\Omega$, the peaks of their thermal freeze-out time distributions are located around 10 ${\rm fm}/c$. While the peaks of $p$ and $\Lambda$ distributions are around 20-30 ${\rm fm}/c$. Due to a large amount of hadronic scatterings during the late {\tt UrQMD} evolution, the thermal freeze-out time distributions of $\pi$ and $K$ widely spread along the time axis. Fig.~3 thus illustrates that thermal freeze-out procedures are hadron species dependent. Compared with other hadrons, the two multi-strange hadrons $\Xi$ and $\Omega$ experience early thermal freeze-out due to their much smaller hadronic cross sections~\cite{SongMultistrange}.


\section{Summary\label{sec:summary}}
\quad In this proceeding, we briefly review our recent investigations on the chemical and thermal freeze-out of various hadron species in 2.76 A TeV Pb+Pb collisions  within the framework of {\tt VISHNU} hybrid model. We found that the chemical and thermal freeze-out procedures are hadron-species dependent.  Due to their much smaller hadronic cross sections, the two multi-strange hadrons $\Xi$ and $\Omega$ experience earlier chemical and thermal freeze-out than other hadrons. To extract the possible effective chemical freeze-out temperatures of various hadron species requires detailed analysis of the space-time distributions of the last inelastic collisions, which we would like to leave it to future study.

\section*{Acknowledgments}
\quad This work was supported by the NSFC and the
MOST under Grants No. 11435001 and No. 2015CB856900 and the China Postdoctoral Science Foundation under Grant No. 2015M570879. We gratefully acknowledge extensive computing resources provided to us on Tianhe-1A by the National Supercomputing Center in Tianjin, China.

\end{document}